\documentclass[12pt]{article}
\pdfoutput=1

\usepackage{amssymb,amsmath,bm}
\usepackage{cite}
\usepackage{graphicx}
\usepackage{longtable,color}

\def\simge{\mathrel{\rlap{\raise 0.511ex \hbox{$>$}}{\lower 0.511ex \hbox{$\sim$}}}}
\def\simle{\mathrel{\rlap{\raise 0.511ex \hbox{$<$}}{\lower 0.511ex \hbox{$\sim$}}}}
\def\slash#1{\setbox0=\hbox{$#1$}\dimen0=\wd0
      \setbox1=\hbox{/} \dimen1=\wd1 \ifdim\dimen0>\dimen1
      \rlap{\hbox to \dimen0{\hfil/\hfil}} #1                        \else
      \rlap{\hbox to \dimen1{\hfil$#1$\hfil}}
      /   \fi}

\newcommand{\lsim}{
\mathrel{\hbox{\rlap{\hbox{\lower4pt\hbox{$\sim$}}}\hbox{$<$}}}}

\newcommand{\gsim}{
\mathrel{\hbox{\rlap{\hbox{\lower4pt\hbox{$\sim$}}}\hbox{$>$}}}}

\allowdisplaybreaks[2]

\newcommand{\be}{\begin{equation}}
\newcommand{\ee}{\end{equation}}
\newcommand{\bea}{\begin{eqnarray}}
\newcommand{\eea}{\end{eqnarray}}

\newcommand{\bi}{\begin{itemize}}
\newcommand{\ei}{\end{itemize}}

\begin{document}
\begin{titlepage}
\vspace*{-1.0truecm}

\begin{flushright}
TUM-HEP-771/10\\
\end{flushright}

\vspace{0.6truecm}

\begin{center}
\boldmath

{\Large\textbf{Characterising New Physics Models by
\\[0.2cm] Effective Dimensionality of Parameter Space }}

\unboldmath
\end{center}

\vspace{0.4truecm}

\begin{center}
{\bf Thorsten Feldmann, Christoph Promberger and Stefan Recksiegel
}
\vspace{0.4truecm}

{\footnotesize
{\sl Physik Department, Technische Universit\"at M\"unchen,
James-Franck-Stra{\ss}e, \\D-85748 Garching, Germany}\vspace{0.2truecm}

}

\end{center}

\vspace{0.5cm}
\begin{abstract}
\noindent

We show that the dimension of the geometric shape formed by
the phenomenologically valid points inside a multi-dimensional parameter 
space can be used to characterise different new physics models
and to define a quantitative measure for the distribution of
the points. We explain a simple algorithm
to determine the box-counting dimension from a given set of parameter points,
and illustrate our method with examples from different models 
that have recently been studied with respect to precision flavour observables.

\end{abstract}

\end{titlepage}
\setcounter{page}{1}
\pagenumbering{arabic}

\section{Introduction} \label{sec:intro}

With the LHC era having started, we are eagerly awaiting new particles
or interactions to be detected at one of the dedicated high-$p_T$ experiments.
Many of the alternative New Physics (NP) models under consideration introduce
a set of additional parameters, notably in the flavour sector, where new
sources for flavour transitions are introduced in the presence of
additional fermionic or bosonic matter fields. 
The multi-dimensional parameter space of NP models
is already constrained by phenomenological data: On the one hand, this
implies exclusion limits, e.g.\ lower bounds on the masses of new particles.
On the other hand, certain parameters or parameter combinations are less
constrained, leaving room for more or less pronounced deviations from the
Standard Model (SM). Depending on the particular NP models, the NP effects
on different observables also show certain patterns of correlations,
which can, for instance, be identified on the basis of a big sample
of theoretically and phenomenologically allowed (but otherwise random) 
parameter points. The interplay between the
results from direct production of new particles at the LHC and from 
the indirect constraints on flavour parameters in the quark and lepton sector 
will play a crucial role for establishing/excluding physics beyond the SM 
in the LHC era, see e.g.~\cite{delAguila:2008iz,Buchalla:2008jp,Raidal:2008jk}
and references therein.

In this paper, we will focus on the multi-dimensional parameter space 
of NP models and its interpretation, after the phenomenological constraints
have been implemented in order to identify valid parameter points. 
A popular method to visualise correlations is to generate scatter plots 
for one observable or parameter against the other, on the basis of these points.
The drawback of such a method is that only low-dimensional projections of
parameter space can be studied. Moreover, the number or number density of points
in certain regions of the scatter plots does not have an immediate statistical
meaning. Below, we will advocate an alternative way to characterise
the space of valid parameter points in a given NP model by its box-counting
dimension (BCD): It takes into account the complete multi-dimensional structure 
of parameter space and is independent of the number of generated parameter 
points. In the next section we will first explain how the dimensionality of parameter
space can be determined using the box-counting algorithm, which will be illustrated
for some simple examples. Our method will then
be applied to two explicit examples for NP models confronted with flavour phenomenology,
and we will show how the BCD of parameter space can be used to classify
different NP scenarios.

\section{The box-counting algorithm}

The BCD (or Minkowski dimension) of a set $S$ in a Euclidean
space ${\cal R}^n$ is defined as
\begin{equation} \label{dimdef}
d={\rm dim}_{\rm box}(S)=\lim_{\epsilon\to 0}{\log N(\epsilon)\over\log 1/\epsilon}
\end{equation}
where $N(\epsilon)$ is number of boxes of side length $\epsilon$ needed to
cover $S$. For the cases that we consider, the BCD is
equivalent to the Hausdorff dimension \cite{Hausdorff:1919}, which is frequently used to describe
fractals.

In practise, we find the BCD by subdividing the parameter space
into $(2^i)^n$ boxes ($n$ is the dimension of the Euclidean space, i.e.\ the
total number of parameters) and plotting the logarithm (base 2) of the 
fill ratio $f$ against $i$. For very small $i$, all boxes will be filled. For
large $i$, if $(2^i)^n \gg p$ (where $p$ is the number of valid points in
parameter space that we found), no box will contain more than one point
and the fill ratio will have a linear slope $-n$ in the logarithmic plot:
\begin{equation}
\log_2 f(i+1) = \log_2 {p\over (2^{i+1})^n}=\log_2 {p\over (2^i)^n}-n=\log_2 f(i)-n
\end{equation}
For intermediate values of $i$, the slope in the logarithmic plot gives us the dimension $d$
as $(d-n)$:
From (\ref{dimdef}), with $\epsilon=1/2^i$ and $f(i)=N(1/2^i)/(2^i)^n$,
\begin{equation} \label{dfrombc}
d={\log_2 f(i)\cdot (2^i)^n \over \log_2 2^i} \quad \Rightarrow \quad
\log_2 f(i+1)-\log_2 f(i)=d-n \,.
\end{equation}

\subsection{Example: The BCD of the coastline of Britain}
The box counting algorithm can be used to show that the
western coastline of Britain has a dimension of 
$d=1.25$\cite{Mandelbrot:1967}.
\begin{figure}[htbp]
\begin{center}
\includegraphics[width=.55\textwidth]{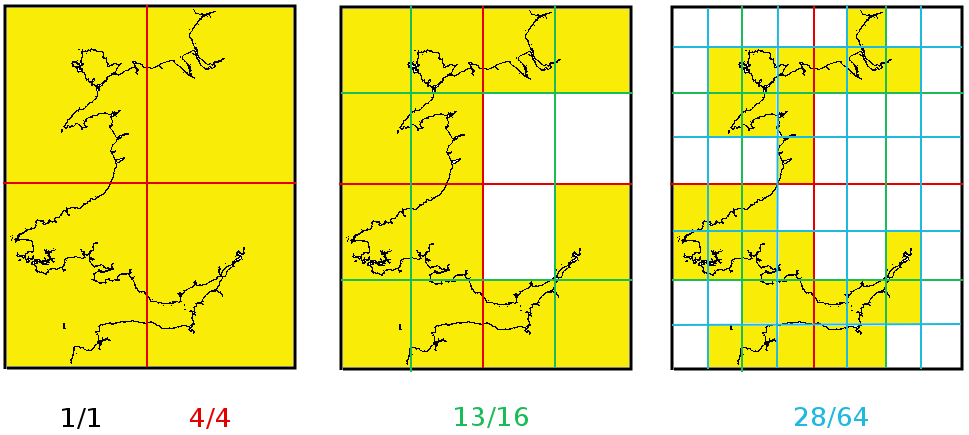}\hspace{.04\textwidth}
\includegraphics[width=.4\textwidth]{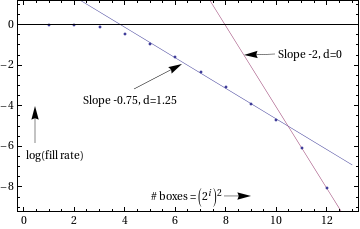}
\end{center}
\caption{The dimension of the coastline of Wales determined by box counting
\label{boxcounting}}
\end{figure}
As shown in Fig.~\ref{boxcounting}, the fill rate of successively
smaller boxes will continually decrease. From the slope of the
curve in the logarithmic plot, we can read off $d=2-0.75=1.25$, according to
Eq.~\ref{dfrombc}. For large $i$, the slope changes to $-2$,
at this resolution we probe the single pixels of our image
that have a dimension of $2-2=0$.

It is also possible to find an approximate self-consistency relation
between the total number of pixels, $N_{\rm tot}$, the extracted BCD, $d$,
and the iteration $i_0$ where the pixel resolution becomes too small and
the kink in Fig.~\ref{boxcounting} occurs,
\begin{align} \label{crit}
 i_0 \, d & \simeq \log_2 N_{\rm tot} \,.
\end{align}
For the above example, we started with $N_{\rm tot} = 15600$ non-white pixels, 
which for $d=1.25$ corresponds to $i_0 \simeq 11$. This is in perfect agreement
with the kink position read off from Fig.~\ref{boxcounting}.

\subsection{Example: Solutions of {\boldmath $f(x)=\sin(1/x)$} }

\label{sec:toy}

\begin{figure}[thb]
 \includegraphics[width=0.45\textwidth]{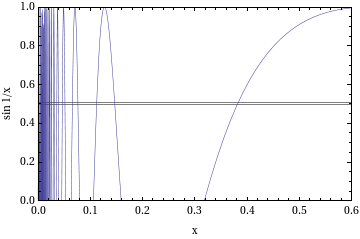} \qquad
 \includegraphics[width=0.45\textwidth]{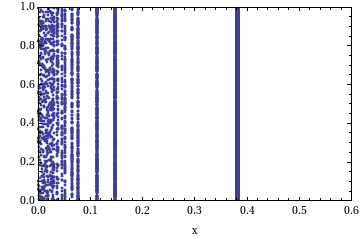}
\caption{\label{fig:toy1} Left hand side: Plot of $\sin 1/x$ indicating
the range of $O(x) = 0.5 \pm \Delta O$ for $\Delta O=0.5\%$.
Right hand side: Geometric shape of the corresponding solutions $\{x(O)\}$.}
\end{figure}

\begin{figure}[thb]
 \includegraphics[width=0.45\textwidth]{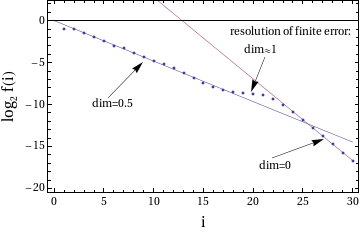} \qquad
 \includegraphics[width=0.45\textwidth]{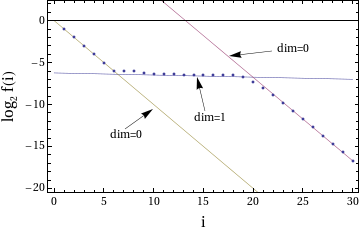}
\caption{\label{fig:toy2} Left hand side: Logarithm of the fill ratio
for $O(x) =\sin 1/x$, Right hand side: Logarithm of the fill ratio
for $O(x) =\sin x$.} 
\end{figure}

\begin{figure}[thb]
 \includegraphics[width=0.245\textwidth]{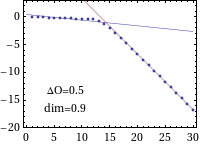}
 \includegraphics[width=0.245\textwidth]{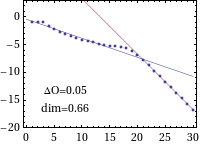}
 \includegraphics[width=0.245\textwidth]{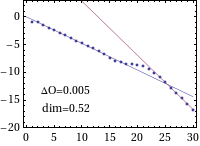}
 \includegraphics[width=0.245\textwidth]{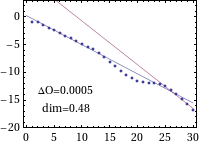}
\caption{\label{fig:varyDeltaO} BCD for different values of $\Delta O$.}
\end{figure}

For later discussion it is useful to consider another toy example, where
we imagine that some physical observable $O$ depends on a fundamental
theory parameter $x$ through
\begin{align} \label{toyconstr}
 O(x) & =\sin(1/x) \,.
\end{align}
Having measured $O$ with some resolution $\Delta O$, 
we may ask for random points $x$ satisfying (\ref{toyconstr})
within the uncertainties.
Because of the non-trivial behaviour of $O(x)$, for a finite
uncertainty $\Delta O$, the geometric
shape of the resulting scatter plot (see Fig.~\ref{fig:toy1})
will correspond to a non-integer BCD. 
In Fig.~\ref{fig:toy2} we show how the behaviour of  $O(x)=\sin(1/x)$
differs from that of $\sin x$. While for $\sin(1/x)$ we observe the
non-integer BCD, for $\sin x$ we first have $d=0$ (only one $x$ value
reproduces the chosen $O(x)$), then $d=1$ (when we resolve the
thickness of the chosen $\Delta O$, and then $d=0$ again (when
we resolve the individual points).
As expected, decreasing the uncertainty $\Delta O$ also leads to a decreasing BCD,
and in the limit $\Delta O \to 0$, one finds
some finite value corresponding to the dimensionality 
of the space formed by the set of solutions $\{x(O) \}$
in the considered interval for $x$,
see (Fig.~\ref{fig:varyDeltaO}).
We would like to stress that in this toy model, 
we do \emph{not} measure a fractal dimension since $\sin (1/x)$ 
is not self-similar. Still, \emph{for a given starting box}, the BCD
provides a useful quantitative measure of the distribution of
the valid data points.

Let us for completeness compare this with the 
traditional measure for fine-tuning,
derived from the normalised derivative \cite{Barbieri:1987fn},
\begin{align}
 \delta_x  \equiv \left| \frac{x}{O(x)} \, \frac{dO(x)}{dx} \right| &= \left| \frac{\cot(1/x)}{x} \right|   \,,
\label{fine}
\end{align}
which can be calculated for every valid parameter point $x(O \pm \Delta O)$. 
We find that the $\delta_x$ calculated in this way is different for
each of the strips contributing to $\{x(O)\}$ and therefore not a
good measure for  the \emph{global} distribution of points  
in this case, as we do not know how to form a meaningful average.

\section{Application to New Physics Models}
\label{results}

In this section we apply our method to two examples of new physics models,
where the valid parameter space is described in terms of a set of points
fulfilling the known constraints from flavour (and if applicable electroweak)
observables.

\subsection{A sequential fourth generation}

One of the simplest extensions of the Standard Model (SM) is
the addition of a sequential fourth generation (4G) of quarks and leptons 
\cite{Frampton:1999xi} (for recent work, see e.g.\ 
\cite{Hou:2005yb,Hou:2006mx,Soni:2008bc,Soni:2010xh,Bobrowski:2009ng,Eberhardt:2010bm,Buras:2010pi,Buras:2010nd,Buras:2010cp}).
We have studied this model extensively
in \cite{Buras:2010pi}, focusing on the phenomenological constraints
and implications in the quark sector.
We will use the parameter points generated
for that paper in the subsequent analysis.

In the 4G quark sector, we have 10 parameters, of which 4 are SM parameters: The usual
three mixing angles and the CKM phase. 
We perform two separate analyses of the 4G parameter space,
either considering the space of all 10 parameters, 
or focusing only on the six new parameters.
In both cases, we have to specify the multi-dimensional starting box,
i.e.\ the lengths of the sides along the directions corresponding to
the individual theoretical parameters. As a standard reference, we consider 
\begin{align}
& \mbox{300~GeV} \leq m_{t'} \leq \mbox{600~GeV} \,, 
&  0 \leq \theta_{ij} \leq \pi/2 \,, \qquad 
   0 \leq \delta_{ij} \leq 2\pi \,.
\end{align}

In this case, we find that -- just as for the example of the coastline of Britain --
there is a clearly defined linear region with a slope larger
than $-n$ in the logarithmic plot of the fill rate, before the
curve bends down to $-n$ for larger $i$.
\begin{figure}[htbp]
\begin{center}
\includegraphics[width=.5\textwidth]{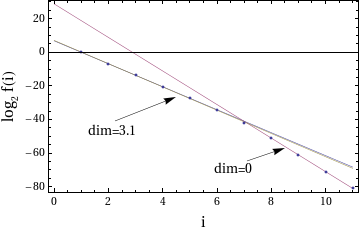}
\end{center}
\caption{Logarithmic plot of the fill rate for the 4G model, considering 
the 10-dimensional SM+NP parameter space. The steeper curve corresponds to
a linear fit to the asymptotic behaviour ($i>6$), whereas the two other
curves denote linear fits for $2\leq i \leq 5(6)$, with the slopes determining
$d_{4G}-10$.
\label{logplot4G}}
\end{figure}
From the first linear region, we can read off a 
BCD of $d_{\rm 4G}=3.1$ with an uncertainty of approximately ${} \pm 0.1$.

This result is very stable: It does not depend on the number
of generated points (which we varied between 5000 and $10^6$), and it
also gives the same result, whether we include the SM parameters or not:
Including the SM parameters for a total number of 10 parameters,
the slope of the curve is approx.\ $-6.9$ (yielding the result
$d_{\rm 4G}=3.1$ cited above). Analysing only the NP parameters, the
number of parameters drops to 6, but the slope of the
curve changes to $-3.0$, giving us the almost identical result of
$d_{4G}=6.0-3.0=3.0$.
\begin{figure}[htbp]
\begin{center}
\includegraphics[width=.5\textwidth]{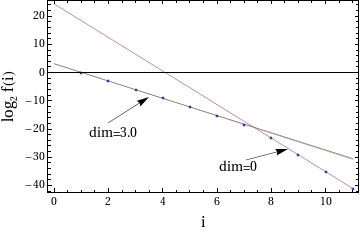}
\end{center}
\caption{Logarithmic plot of the fill rate for the 4G model, considering 
only the 6-dimensional NP parameter space. The steeper curve corresponds to
a linear fit to the asymptotic behaviour ($i>6$), whereas the two other
curves denote linear fits for $2\leq i \leq 5(6)$, with the slopes determining
$d_{4G}-6$.
\label{logplot4G-NPonly}}
\end{figure}

%
%
%
We stress here that the numerical result for the BCD
depends on the size of the chosen starting box (cf.\ the toy example
in Sec.~\ref{sec:toy}). It can be
seen that this indeed needs to be the case by considering
the limiting cases of very small or very large boxes. Boxes
smaller than the range of valid parameters are problematic
for objects that are not completely self-similar (e.g.\ 
because of limited resolution: Looking only at the bay of
Swansea will give a result different from that for the
complete coastline of Wales).
If, on the other hand, the box size is much larger in
some dimension than the parameter range allowed for the
corresponding parameter, the variation in that parameter
cannot be resolved and will not contribute to the effective
dimension of the parameter space (on a scale of $10^9$km,
the coastline of Wales has no extent).
In between these two extreme cases, we
expect a residual logarithmic dependence on the size of the starting box.

\begin{figure}[htbp]
\begin{center}
\includegraphics[width=.5\textwidth]{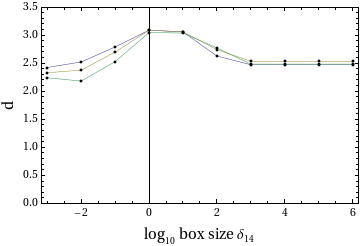}
\end{center}
\caption{Plot of the effective dimension against the logarithm
of the box size for the parameter $\delta_{14}$. The three
lines correspond to the fits for $1,2,3\le i \le 6$.
\label{logplot4G-varyrange}}
\end{figure}
This is in fact the case as shown in Fig.~\ref{logplot4G-varyrange}
where the range for the 4G parameter $\delta_{14}$ is varied:
For ranges from $10^{-3}$ to $10^0$, the granularity of our data points
shows and the determined dimension is too small. Between 
$10^0=1$ and $10^1=10$ (where the {\em natural} value of $2\pi$ lies),
we obtain our result of $\sim 2.9$, and between $10^1$ and $10^3$ we
see the predicted fall-off. From $10^3$ on, the box is so large that
the variation of $\delta_{14}$ cannot be resolved anymore and we
obtain the dimension produced by the other parameters. At the same
time we can see that the contribution of $\delta_{14}$ to the effective
dimension of the whole parameter space is about $0.5$.

\subsubsection{Ranking of individual 4G scenarios}

As has been discussed in \cite{Buras:2010pi}, the 4G parameter space
can be divided into different regions which are characterised by the
scaling of the 4G mixing angles with the Wolfenstein parameter $\lambda \simeq 0.22$,
\begin{align}
 \left(\theta_{14},\theta_{24},\theta_{34}\right) & \sim \left(\lambda^{n_1},\lambda^{n_2},\lambda^{n_3}\right) \,.
\end{align}
Each triple of integers $(n_1,n_2,n_3)$ then defines an
individual 4G scenario. It has already been seen in \cite{Buras:2010pi}
that different scenarios lead to rather different correlations, both
between physical observables and between the new 4G CP phases.

We may ask ourselves whether this behaviour is also reflected
in the BCD for the (now further restricted) parameter space within 
a given scenario $(n_1,n_2,n_3)$. Indeed, we find that the 
dimension $d_{n_1n_2n_3}$ 
of individual scenarios is quite distinct. 
We can thus define a ranking between different scenarios according
to the value of $d_{n_1n_2n_3}$. In Table~\ref{tab:rank4G}, we show
such a ranking for a selection of ``interesting'' scenarios 
which have been identified in \cite{Buras:2010pi}.
\begin{table}[tpbh]
 \caption{\label{tab:rank4G} Ranking of some individual 4G scenarios according
to their BCD.}
\begin{center}
 \begin{tabular}{|c|c|c||}
\hline 
 Scenario ($n_1,n_2,n_3$) & Dimension $d_{n_1n_2n_3}$ & \# of Points
\cr 
\hline\hline
211 & 1.7 & 5486
\cr 
221 & 1.7 & 5084
\cr 
\hline 
231 & 2.1 & 37k
\cr 
432 & 2.2--2.4 & 29k
\cr 
431 & 2.4--2.5 & 56k
\cr 
332 & 2.5--2.6 & 125k
\cr 
331 & 2.5 & 206k
\cr 
321 & 2.6 & 295k
\cr 
\hline\hline  
 \end{tabular}

\end{center}

\end{table}
We have also quoted the number of points that have been found 
by the numerical procedure in \cite{Buras:2010pi} for
the corresponding region of parameter space. Notice that the
number or density of valid points does not necessarily
allow for a quantitative or qualitative interpretation of
the NP model under consideration, as it usually depends on
the way the parameter points are generated (although in the
shown case, the numerical procedure has treated all parameter
values on equal footing, and therefore a smaller number of points
also corresponds to a smaller BCD).

We observe that the scenarios 211 and 221 have the smallest dimension,
with a value significantly below 2 and thus more than 1 unit lower than
for the total valid parameter space. This is in line with the findings in
\cite{Buras:2010pi}, where these scenarios have been shown to give
the most stringent constraints on the new 4G phases $\delta_{14}$ and $\delta_{24}$,
and thus effectively reducing the dimensionality of the accessible parameter space
by 1--2 units.
(At the same time, these scenarios predict the most drastic and interesting
deviations of flavour observables from their SM values.)

\subsubsection{BCD of ``forbidden'' 4G scenarios}

The 4G model provides a particular realisation of next-to-minimal
flavour violation, according to the definition in \cite{Feldmann:2006jk}:
It contains new sources of flavour and CP violation; still the new mixing
angles are not expected to be generic but --- as in the examples discussed
in the previous subsection --- naturally feature similar hierarchies as known from
the 3G SM. Moreover, certain scenarios $(n_1,n_2,n_3)$ 
can be (formally) excluded because of self-consistency inequalities among 3G and 4G mixing angles \cite{Feldmann:2006jk,Buras:2010pi},
\begin{align} \label{eq:4Gforbid}
 \theta_{ik} \theta_{jk} \lesssim \theta_{ij} \qquad \mbox{(\small $i,j,k=1\ldots 4$, no summation over $k$).}
\end{align}
If these inequalities are violated, we expect a certain amount of fine-tuning
between 4G parameters in order to keep the off-diagonal elements in the 4G
mixing matrix sufficiently small.

Again, the BCD for such scenarios provides
a measure to quantify this effect. In Table~\ref{tab:4Gforbid} we 
compare different scenarios that are classified by their minimal distance 
$\Delta^2= \sum_{i=1}^3 \left(\Delta n_i\right)^2$ from one of the
allowed scenarios (such that the allowed scenarios correspond to $\Delta^2=0$),
For the forbidden scenarios, with increasing $\Delta^2$, we expect more and
more fine-tuning and thus a smaller BCD $d_\Delta$. This is clearly
confirmed by the numerical analysis which, for scenarios close to
the allowed regions (i.e.\ $\Delta^2=1$) yields a BCD close to the one
found from the allowed scenarios, whereas very distant scenarios yield
BCDs which are more than one unit smaller.

\begin{table}[tpbh]
 \caption{\label{tab:4Gforbid} BCD of ``allowed`` and ``forbidden'' 4G scenarios
for different distance $\Delta^2$.}
\begin{center}
 \begin{tabular}{|l c|c|c||}
\hline 
 Scenarios ($n_1,n_2,n_3$)&  Distance $\Delta^2$& Dimension $d_{\Delta}$ & \# of Points
\cr 
\hline\hline
(3,2,1), (3,2,2), (3,3,1), (3,3,2), \ldots & 0 & 2.60--2.80 & 166k
\cr 
\hline \hline 
(3,4,1), (4,2,1), (5,3,1), (4,4,1), \ldots & 1 & 2.55--2.75 & 150k
\cr 
\hline
(5,2,1), (6,3,2), (3,2,6), (2,1,5), \ldots & 2 &  1.9--2.0 & 7.8k
\cr 
\hline
(6,3,1), (2,5,1), \ldots & 3 & 1.8--1.9 & 4.5k
\cr 
\hline 
rest & $\geq 5$ & 1.2--1.3 & 354
\cr 
\hline\hline 
\end{tabular}
\end{center}
\end{table}

\subsection{The Littlest Higgs Model with T parity}

An elegant solution to the hierarchy problem are the Littlest Higgs
Models \cite{ArkaniHamed:2001ca,ArkaniHamed:2001nc,Schmaltz:2005ky,
Perelstein:2005ka,ArkaniHamed:2002qy}
with $T$--parity (LHT) 
\cite{Cheng:2003ju,Cheng:2004yc,Low:2004xc}
which have been analysed in \cite{Blanke:2006sb,Blanke:2006eb,Blanke:2009am}.
In this model, we encounter 9 new flavour parameters: Three mixing angles and three phases in the
mixing matrix for the mirror quarks, and three mirror quark masses.
The valid points in parameter space are distributed rather evenly 
over the available space: Of the $2^9=512$ boxes for $i=2$,
499 are filled (c.f.\ 8/1024 in the 4G case).
For larger $i$, the fill rate quickly falls with $(2^i)^9$
as expected for a dimensionless object.
Unlike in the 4G model, we do not observe a kink corresponding
to a non-trivial BCD, irrespective of the chosen starting values
for the bounding box. In principle, this result allows for two
different interpretations:
\begin{enumerate}
 \item The dimensionality of the theoretical parameter space is indeed compatible with $d \simeq  0$.
Indeed, this could have been anticipated from the results in \cite{Blanke:2006sb,Blanke:2006eb,Blanke:2009am},
where these points have been characterised 
by the Barbieri--Giudice fine-tuning measure
\cite{Barbieri:1987fn}, cf.\ Eq.~(\ref{fine}), yielding widely varying
values up to and exceeding ${\cal O}(100)$.
\item Due to the limited number of generated valid parameter points for the LHT model
 (which, in turn, can be understood as a consequence of the required fine-tuning,
  respectively the small dimensionality of parameter space), the BCD measurement,
  in principle, allows for a second solution, when $N_{\rm tot}$ is too small to
  resolve the kink at $i_0 \geq 2$ related to the ``true'' dimensionality. 
  With  (\ref{crit}) this can be easily translated to a bound on $d$,
\begin{align}
n \geq d & \gtrsim \log_2 N_{\rm tot} /2 \,. 
\end{align}
 In fact, for the LHT we only consider ${\cal O}(5k)$ points, and therefore, on the basis
 of the box-counting method alone, we cannot
 exclude BCDs between $6 \lesssim d < 9$. This means that the allowed
 points might lie on a structure that is too complex to be resolved with
 the number of points that we have at our disposal.

\end{enumerate}

\begin{figure}[htbp]
\begin{center}
\includegraphics[width=.5\textwidth]{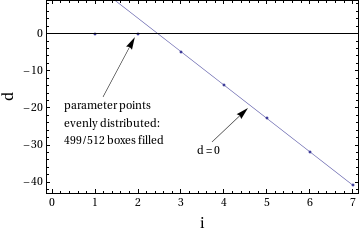}
\end{center}
\caption{Logarithmic plot of the fill rate for the 9-parameter space of the 
LHT model.
\label{logplotLHT}}
\end{figure}

\section{Summary}

We have proposed a novel method to study the distribution of
parameter points in their multi-dimensional space. By employing
a very simple box counting algorithm that can be applied to any
data set, we obtain a measure for the effective number of free
parameters and the amount of correlations induced by the phenomenological
constraints.
Unlike traditional measures of fine-tuning, the BCD
method also works if the valid points have very different
individual fine-tuning. Additionally, the BCD method uses
only the valid data points and does not need to solve 
analytical expressions for the observables.

Using the well known example of the fractal dimension of coastlines
and a toy model, it can be shown that the new measure makes sense
even under variations of the box size and the number of points.
The proposed method gives an easy and quick measure of the
distribution of the points that would otherwise require many
(necessarily low-dimensional) scatter plots to study.

The method is applied to two models of New Physics. For a sequential
fourth generation of quarks, we show that the effective dimension
of the parameter space is $\approx 3$, independent of whether the SM 
parameters are counted as variables or not (which is a highly non-trivial 
observation because of the complex dependence on the SM parameters).
When classifying the points in 4G parameter space into different
scaling scenarios, we find that the corresponding dimensions in
parameter space are (i) smaller than those of the complete data
set (ii) decrease with increasing {\em distance} from {\it allowed}
scenarios.

Comparing the 4G findings with the phenomenologically valid
points in the Littlest Higgs Model with T parity, we find that
in the LHT model the effective dimension is 0 (corresponding to
purely fine-tuned points) or very large (i.e.\ we do not have
enough points to resolve the structure that they lie on).

\subsection*{Acknowledgements}

We thank Bj\"orn Duling and Tillmann Heidsieck for helpful discussions.
This research was partially supported by the Cluster of Excellence `Origin and Structure of the Universe', the Graduiertenkolleg GRK 1054 of DFG and by the German `Bundesministerium f{\"u}r Bildung und Forschung' under contract 05H09WOE.


\bibliographystyle{JHEP}
\bibliography{dimpaper}

\end{document}